# Autonomous push-down automaton built on DNA*

Tadeusz KRASIŃSKI[1], Sebastian SAKOWSKI[1], Tomasz POPŁAWSKI[2]
[1]*Faculty of Mathematics and Computer Science, University of Łódź*
*Banacha 22, 90-238 Łódź, Poland*
[2]*Department of Molecular Genetics, University of Łódź*
*Pomorska 141/143, 90-236 Łódź, Poland*
*e-mail: krasinsk@uni.lodz.pl, sakowski@math.uni.lodz.pl, tplas@biol.uni.lodz.pl*

**Abstract.** In this paper we propose a biomolecular implementation of the push-down automaton (one of theoretical models of computing device with unbounded memory) using DNA molecules. The idea of this improved implementation was inspired by Cavaliere et al. (2005).

**Keywords:** push-down automaton, DNA computing.

## 1. Introduction

In the paper by Cavaliere, Janoska, Yogev, Piran, Keinan, Seeman (2005) the authors proposed a theoretical model (*i.e.* not tested in laboratory) of implementation of the push-down automaton built on DNA. The idea was inspired by two papers: the first one by Rothemund (1995) who proposed a simulation of the Turing machine - the basic theoretical model of the computation - and the second one by Benenson, Paz-Elizur, Adar, Keinan, Livneh, Shapiro (2001) who proposed a simulation of the finite automaton – the simplest model of the computation. The above three implementations represent the all basic theoretical models of computers in the Chomsky hierarchy. But these all simulations have weak points by different reasons.

The Rothemund model is not autonomous. A person must interfere in the process to obtain required sequence of actions of many restrictases. This is likely a reason why nobody tested it experimentally.

It turn, Beneson *et al*. model is autonomous, programmable and was tested in laboratory but it represents the simplest computational model - a finite automaton (in fact it was only 2-states 2-symbols finite automata). The next propositions along the same idea (Soreni *et al.*, 2004; Unold *et al.* 2004; Krasiński and Sakowski 2008) did not improve essentially the situation.

The last Cavaliere *et al.* (2005) model is more-powerful (a push-down automaton), autonomous, programmable (although the action of it was illustrated only on one simple example) but the problem lies in obtaining the right sequence of ligations of transition molecules to the input and to the stack (represented by the same circular DNA). The authors themselves indicate this problem “It is first important to know which side is ligated first, since there is degeneracy in the stack side … and therefore different transition molecules may be ligated at that end at

* This project is supported by the National Science Centre of Poland (NCN). Grant number: DEC-2011/01/B/NZ2/03022.

any stage" and propose two ways to reduce (not eliminate) the problem. Moreover another problem in their model is that it is not clear biochemically whether the used enzyme *Psr*I couldn't cut transition molecules of the first kind (which add the symbol Z to the stack) before ligation it to the input and to the stack.

In the paper we propose an improvement of the last model of push-down automata to avoid these problems. However it is still a theoretical model not tested yet in laboratory. We propose a new shape of transition molecules and another kind of restriction enzymes, which cut only when the ligation of a transition molecule to the circular molecule of the input will be accomplished on both sides.

## 2. Push-down automaton

In this section we recall the definition of the push-down automata (PDA). More information one can find in any textbook (Hopcroft and Ullman, 1979; Sipser, 2006).

A push-down automata is a finite automata (nondeterministic) which has a stack, a kind of simple memory in which it can store information in a last-in-first-out fashion. So PDA has a finite control unit, input tape and a stack.

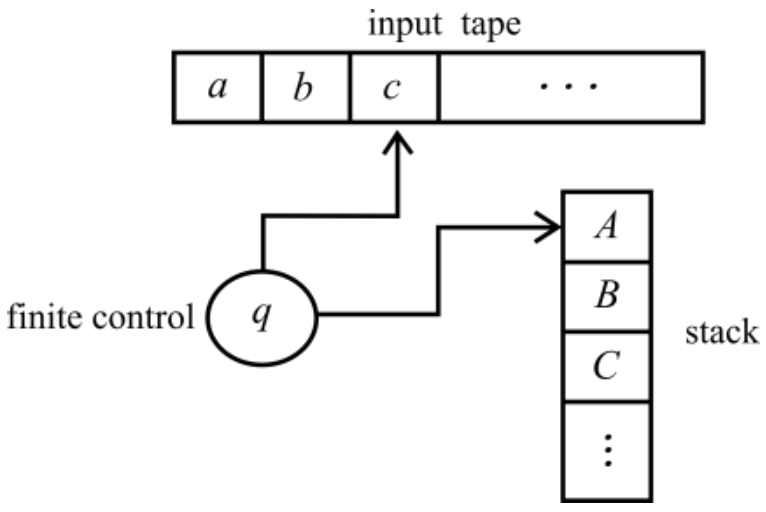

Fig. 1. A scheme of the PDA.

In each step the machine, based on its current state ($q$), the input symbol which is being currently read ($c$) and the top symbol on the stack ($A$) performs a move according to a transition rule (from a list of transition rules associated to a given PDA): pops the top symbol of the stack, push a symbol (or a sequence of symbols) onto the stack, move its read head one cell to the right and enter a new state. We also allow $\varepsilon$ - transitions in which PDA can pop and push without reading the next input symbol. The PDA is nondeterministic, so there may be several transitions that are possible in a given configuration. We will write transition rules in the following way

$$(q,c,A) \to (q',A')$$

where: $q'$ - a new state, $A'$ - a new symbol or a sequence of symbols (may be an empty sequence) which replaces $A$ on the top of the stack.

There are two (equivalent) alternative definitions of acceptance of an input word $w$: by empty stack and by final state. Since in the presented implementation we use the second one we will recall only it. A PDA accept its input word $w$ if it enters a final state (from a distinguished subset of all states) after scanning the entire word *w*, starting from the initial configuration, with *w* on the input tape and with special initial symbol $\perp$ on the stack.

The class of languages accepted by PDA's is the class of context free languages which strictly includes the class of regular languages (accepted by finite state automata) and is strictly contained in the class of recursive enumerable languages (accepted by Turing machines).

*Example* 1. A standard non-regular language accepted by a PDA is

$$L_1 = \{a^n b^n : n \in N\}.$$

A PDA accepting this language has three states: $q_0$ - initial state, $q_1$, $q_2$ - final state. Its transition rules are:

1. $(q_0, a, \perp) \to (q_0, A\perp)$
2. $(q_0, a, A) \to (q_0, AA)$
3. $(q_0, b, A) \to (q_1, \varepsilon)$
4. $(q_1, b, A) \to (q_1, \varepsilon)$
5. $(q_1, \varepsilon, \perp) \to (q_2, \varepsilon)$

*Example* 2. PDA's can add integers. A PDA accepting the language

$$ADD = \{a^n b^m c^{n+m} : n, m \in N\}$$

which will be our main example in implementation has four states: $q_0$ - initial state, $q_1$, $q_2$, $q_3$ - final state. Its transition rules are:

1. $(q_0, a, \perp) \to (q_0, A\perp)$
2. $(q_0, a, A) \to (q_0, AA)$
3. $(q_0, b, A) \to (q_1, AA)$
4. $(q_1, b, A) \to (q_1, AA)$
5. $(q_1, c, A) \to (q_2, \varepsilon)$
6. $(q_2, c, A) \to (q_2, \varepsilon)$
7. $(q_2, \varepsilon, \perp) \to (q_3, \varepsilon)$

A sequence of configurations (state, remaining input word, stack) of this PDA on the input word $aabccc \in ADD$ is:

$$(q_0, aabccc, \perp) \xrightarrow{1} (q_0, abccc, A\perp) \xrightarrow{2} (q_0, bccc, AA\perp) \xrightarrow{3} (q_1, ccc, AAA\perp) \xrightarrow{5}$$

$$(q_2, cc, AA\perp) \xrightarrow{6} (q_2, c, A\perp) \xrightarrow{6} (q_2, \varepsilon, \perp) \xrightarrow{7} (q_3, \varepsilon, \varepsilon)$$ - acceptation,

and on the input word $abc \notin ADD$ is:

$$(q_0, abc, \perp) \xrightarrow{1} (q_0, bc, A\perp) \xrightarrow{3} (q_1, c, AA\perp) \xrightarrow{5} (q_2, \varepsilon, A\perp)$$ - stop the action.

**3. The implementation of PDA**

The implementation of PDA is similar to that of Cavaliere *et al.* (2005) with changes which eliminate their obstacles. The main idea of implementation is as follows.

The basic elements of PDA i. e. the input tape and the stack are represented in the same circular dsDNA molecule which one end represents the stack and the second one the input word (Fig. 2).

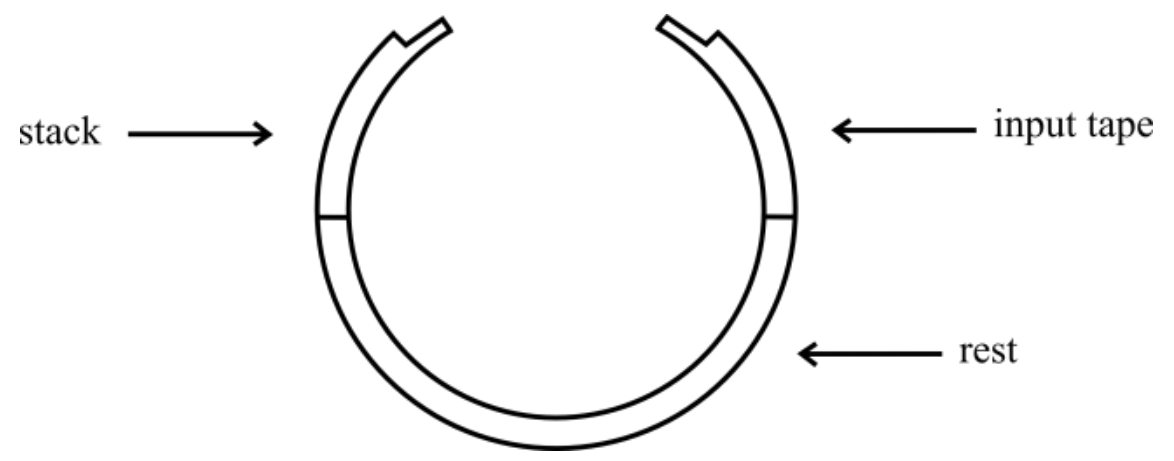

Fig. 2. The basic elements of implementation of PDA.

The sticky end of the stack represents the top symbol on the stack and the sticky end of the input tape represents the first symbol of the input word (to be read) and simultaneously the state of PDA.

The transition rules of PDA are suitable DNA molecules which hybridize to both ends of the circular DNA.

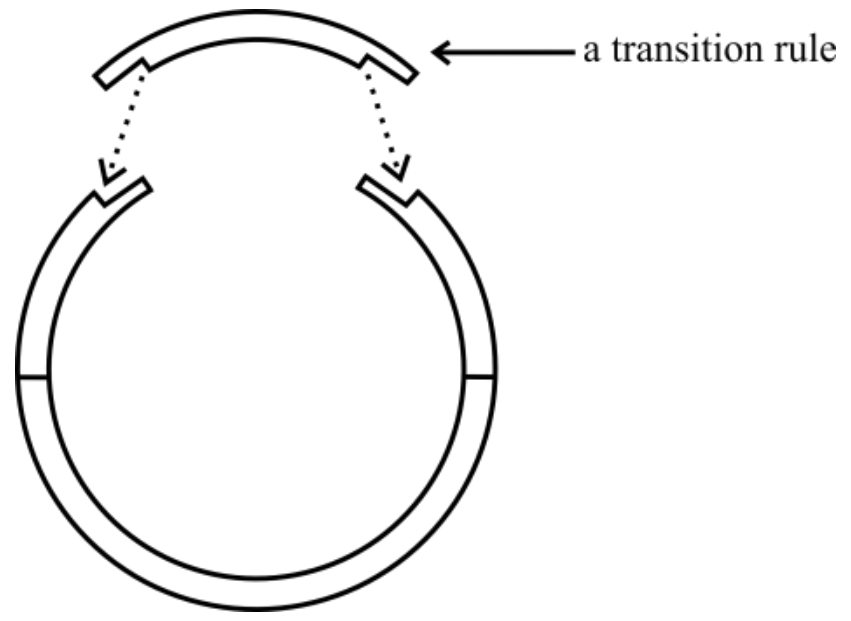

Fig. 3. Process of hybridizing a transition rule to both ends of DNA.

After ligation appropriate restriction enzymes cut this circular molecule. Their actions cause changes in the stack and in the input word according to the move which is represented by this transition molecule. A new idea is that the action of restriction enzymes will take place only when the transition molecule ligate to both ends of the circular molecule. It happens because the chosen restriction enzymes (*Bgl*I) has two separated recognition sites (Fig. 4), which appear both only when a transition molecules ligate to both ends of the circular molecule. After the cut additional molecules and restriction enzymes make adequate changes in the stack and in the input word. Then the next transition rule may act. When a sequence of such transitions lead to reading out the input word and the last sticky end would represent the final state of PDA, then a long additional DNA molecule ligates to the molecule. It can be detected in the solution by gel electrophoresis. The word is accepted.

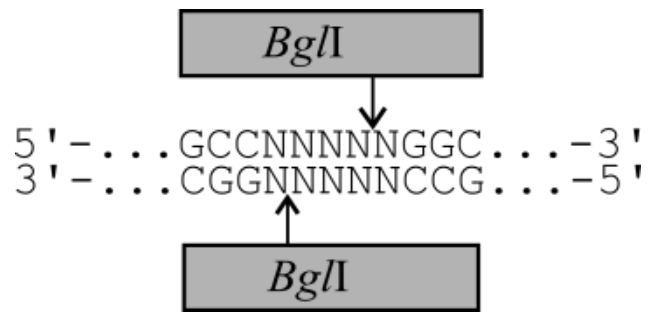

Fig. 4. Acting on the enzyme *Bgl*I.

## 4. The practical implementation

We implement the idea presented in Section 3 on the PDA given in Example 2 (Fig. 5) i.e. on PDA performing addition of integers.

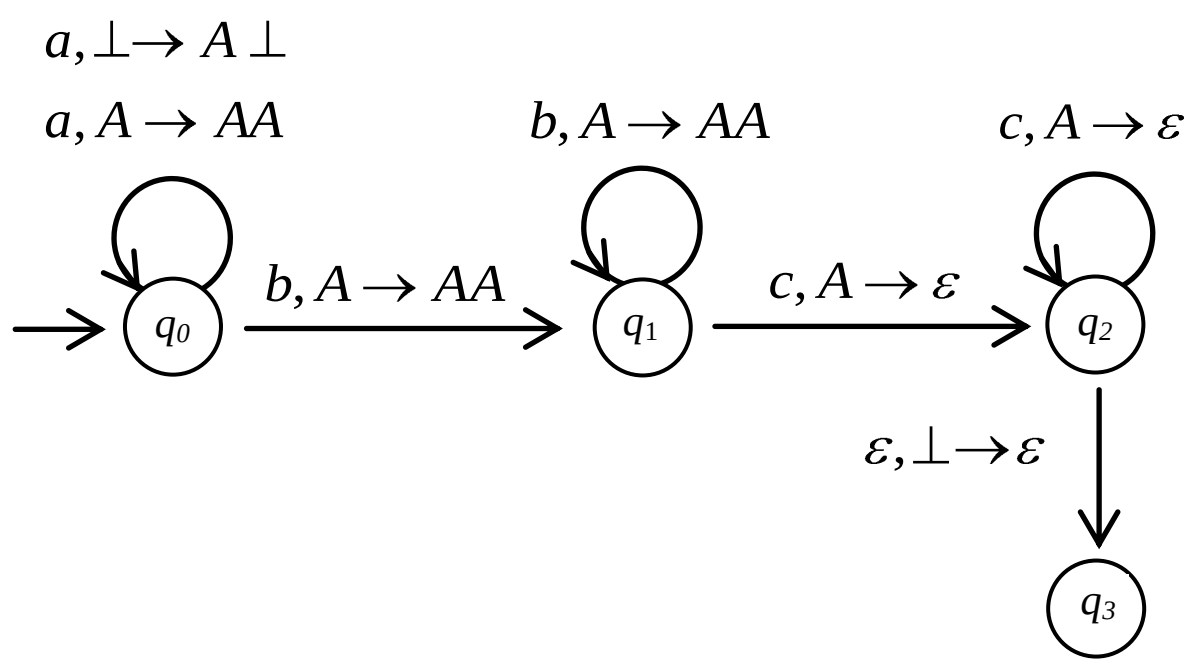

Fig. 5. The graph of a PDA which add integers.

It has seven moves. Each of them is represented by a transition molecule, additional molecules and suitable restriction enzymes, presented in Table 1.

Table 1
The transition rules and their molecular representation

| Transition rule | Transition molecule | Additional molecule | Restriction enzymes |
|---|---|---|---|
| $(q_0, a, \perp) \rightarrow (q_0, A\perp)$ | ATCGCC AGC CG<br>TCG | CTGAAG NNNNN GCC<br>GACTTC NNNNN | *Bgl*I<br>*Acu*I |
| $(q_0, a, A) \rightarrow (q_0, AA)$ | AGC CG<br>CGG TCG | CTGAAG NNNNN GCC<br>GACTTC NNNNN | *Bgl*I<br>*Acu*I |
| $(q_0, b, A) \rightarrow (q_1, AA)$ | AGC CT<br>CGG TCG | CTGAAG NNNN CGG<br>GACTTC NNNN | *Bgl*I<br>*Acu*I |
| $(q_1, b, A) \rightarrow (q_1, AA)$ | AGCC TG<br>CGG TCGG | CTGAAG NNNN GCC<br>GACTTC NNNN | *Bgl*I<br>*Acu*I |
| $(q_1, c, A) \rightarrow (q_2, \varepsilon)$ | AAAG AG<br>CGG TTTC | CTGAAG NNNN AAG<br>GACTTC NNNN<br><br>CGTCG N<br>GCAGC N CTT<br><br>GCTCTTC TCCG<br>CGAGAAG | *Bgl*I<br>*Acu*I<br>*Bbv*I<br>*Sap*I |
| $(q_2, c, A) \rightarrow (q_2, \varepsilon)$ | GGGA AG<br>CGG CCCT | CTGAAG NNNN GGA<br>GACTTC NNNN<br><br>CGTCG N<br>GCAGC N TCC<br><br>GCTCTTC ACCG<br>CGAGAAG | *Bgl*I<br>*Acu*I<br>*Bbv*I<br>*Sap*I |
| $(q_2, \varepsilon, \perp) \rightarrow (q_3, \varepsilon)$ | GGG TA<br>CGG CCC | CTGAAG NNNN GGA<br>GACTTC NNNN<br><br>CGTCG N<br>GCAGC N TCC<br><br>ACCG<br>300 bp<br><br>GG<br>500 bp | *B gl*I<br>*Acu*I<br>*Bbv*I |

The action of the enzyme *Bgl*I is presented in Fig. 4. Remaining enzymes act as follows (Fig. 6).

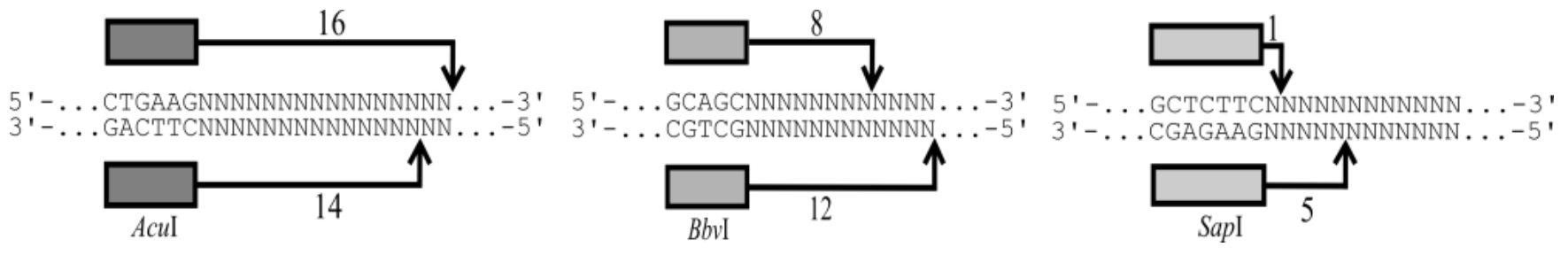

Fig. 6: Action of the enzymes *Acu*I, *Bbv*I, *Sap*I.

The sticky end of an input word represent both - a symbol and a state of PDA according to the rules (Fig. 7).

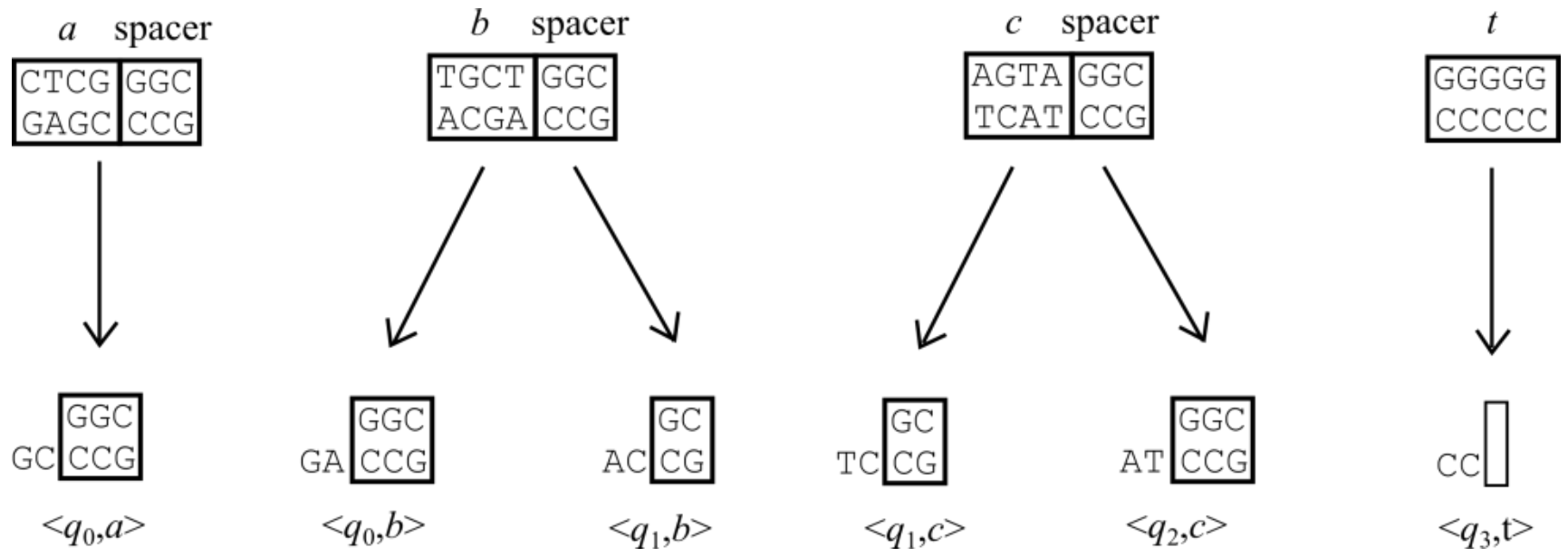

Fig. 7. DNA codes of the symbols and pairs <state, symbol>.

The symbols on the stack $\{A, \perp\}$ and their representations on the top of the stack are (Fig.8).

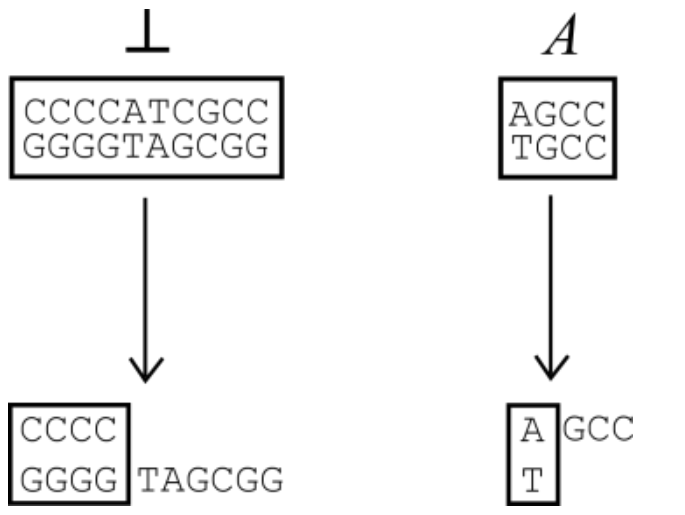

Fig. 8. Representation of the stack symbols.

The representation of the considered PDA with the input word *aabccc* in initial state $q_0$ and the symbol $\perp$ on the stack is shown in Fig. 9.

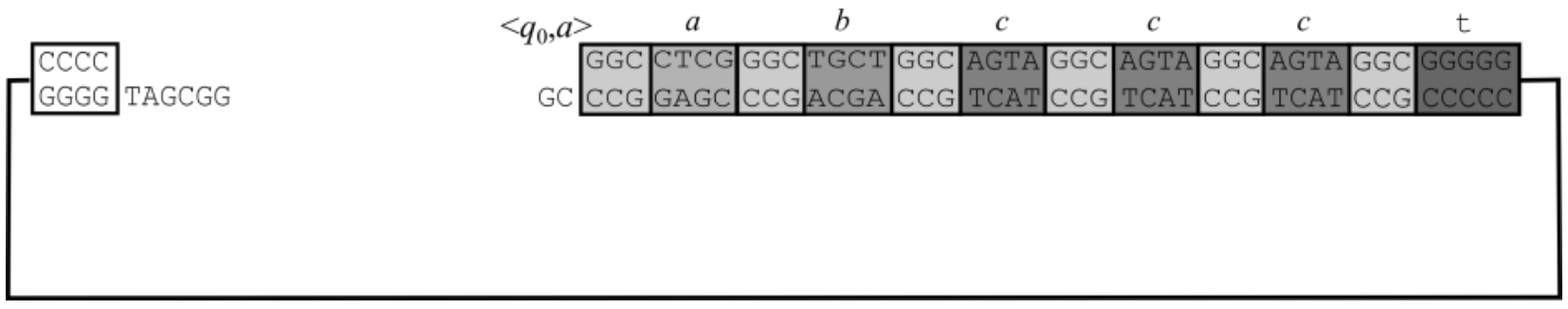

Fig. 9. PDA with the input word *aabccc*.

The action of the PDA will be illustrated on two moves, the first which push a symbol on the stack (Fig. 10) and the second one which pop the symbol of the stack (Fig. 11.).

The main idea of the first move $(q_0, a, \perp) \to (q_0, A\perp)$ which push the symbol *A* on the stack is to use the restriction enzyme *Bgl*I, which cut DNA strand only when transition molecule merge stack and the input tape.

It is caused by the fact that the enzyme *Bgl*I has two separated recognition sites 5'...GCC5(nt)GGC...3' which appear when the transition molecule ligates to the stack and to the input word. An important fact is in using spacers GGC between symbols of the input word. After the cut the second restriction enzyme *Acu*I together with an additional molecule make a change in the input word.

A second move $(q_1, c, A) \to (q_2, \varepsilon)$ which pop the symbol from the stack acts by using restriction enzyme *Bgl*I (Fig. 11). After cutting with the enzyme *Bgl*I we have to remove actual symbols from the input word and the stack. The operation of removing from the input word is the same as in the first move (using the restriction enzyme *Acu*I). Since we couldn't find a commercial enzyme which cut a DNA molecule in a long distance from the recognition site and remain 3-nt sticky end we have to apply two restriction enzymes (*Bbv*I and *Sap*I)

The remaining moves act similarly. The whole process over the word *aabccc* is presented in appendix.

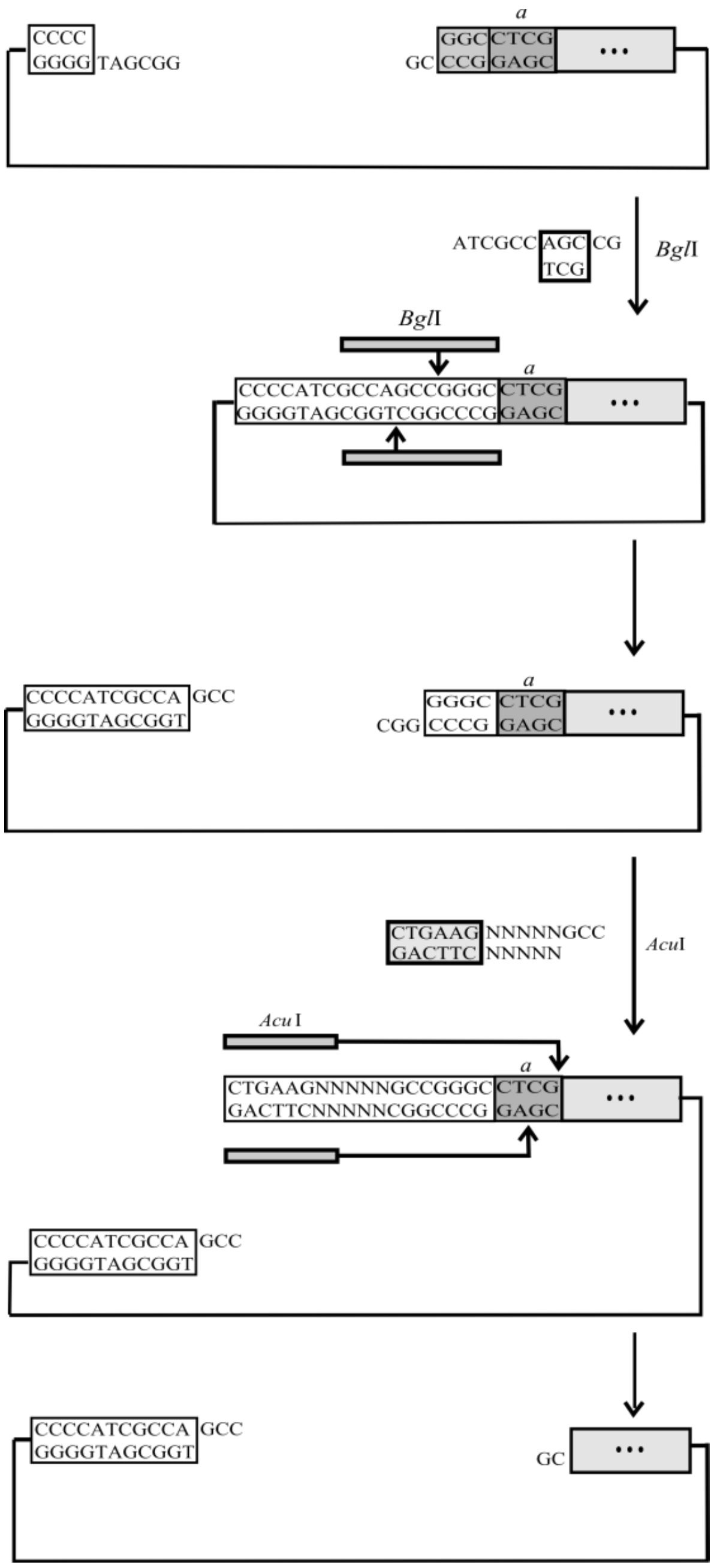

Fig.10. The push a symbol on the stack $(q_0, a, \perp) \rightarrow (q_0, A\perp)$.

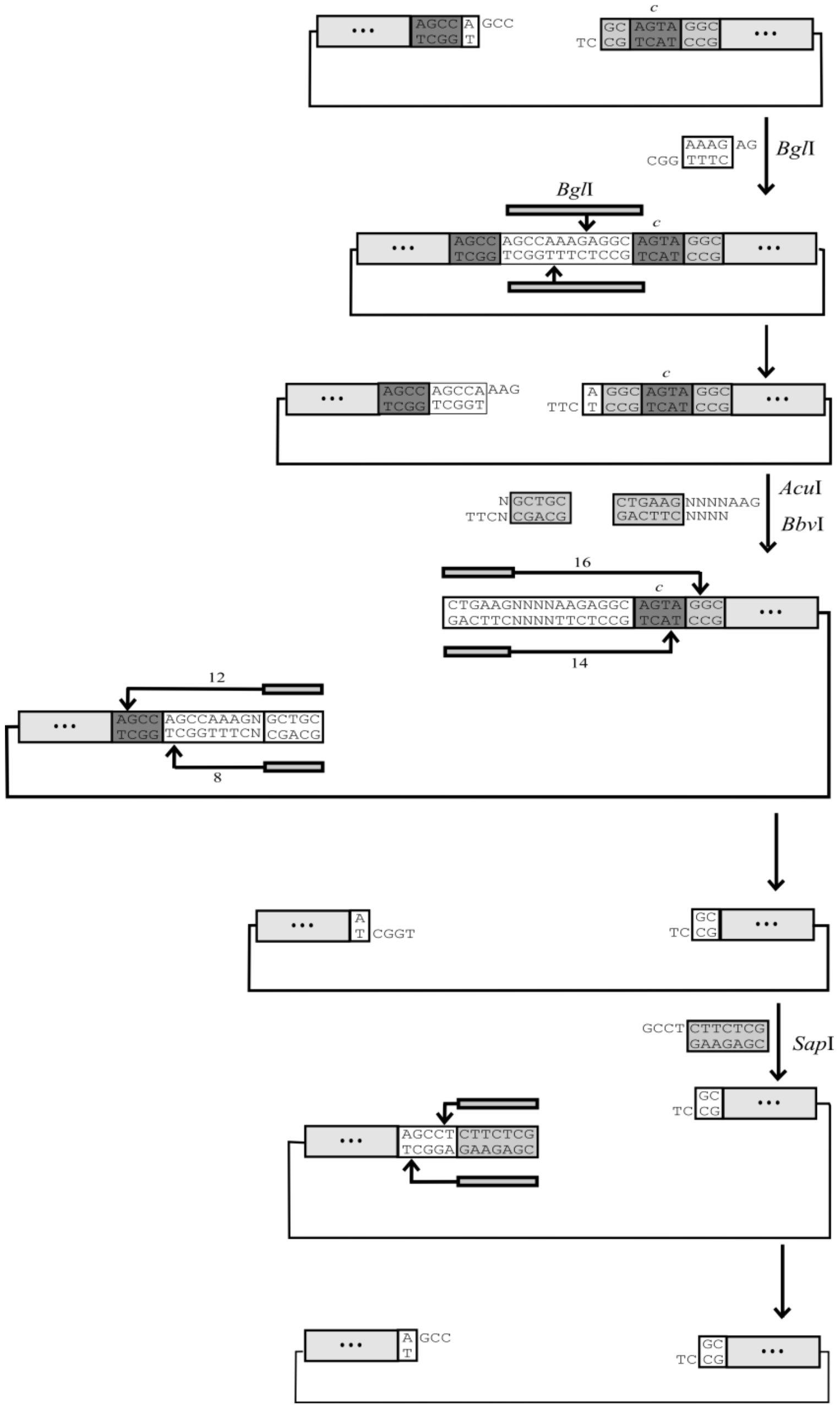

Fig 11. The pop a symbol from the stack $(q_1, c, A) \rightarrow (q_2, \varepsilon)$ .

## 4. Conclusions and Future Work

We have presented a new way to implement push-down automaton based on DNA molecules and restriction enzymes. It is an improved version of the idea presented in Cavaliere *et al*. (2005). Another attempts (not fully matured and functioned) are in Shi *et al*. (2005), Zhang *et al*. (2006), Zhang *et al*. (2008). A new idea is to use a restriction enzyme which have two separated recognition sites. It allows to cut DNA molecules representing elements of a PDA after ligating transition molecules to both sides of circular DNA. It avoids problems appeared in Cavaliere *et al*. (2005). This will enable us in the future to construct more powerful automata than PDA's, which gives possibility to solve more complicated problems. Actually we implemented our theoretical model of finite automata (more powerful than presented in Benenson *et al*. (2001) in laboratory in cooperation with a research group of Department of Molecular Genetics (the University of Lodz). This attempt of laboratory implementation of our research groups is described in Błasiak, Krasiński, Sakowski, Popławski (2011). We tested in laboratory simultaneous action of two restriction enzymes *Acu*I and *Bbv*I which is a crucial step in experiment presented in this paper. The next step could be laboratory implementation of PDA presented in this article.

Circular molecule dsDNA used in our model open new possibility to insert and apply our automaton to the bacteria cell. Such type of DNA molecules are plasmids - heritable DNA molecules that are transmissible between bacterial cells and bacterial genomes. Bacteria controls DNA replication process via origin replication elements. These genetic elements are built with blocks of repeated sequences and replication is initiated when special proteins (e. g. DnaA in *E. coli*) binds to series of repeats. Regulations of bacterial genome and plasmid propagation is possible with use of our automaton by controlling the number of repeat motifs presented in origin (by inserting to the stock or removing from the stock). In similar way it is possible to control in bacteria not only DNA replication but also transcription of some bacterial genes. Transcription starts when RNA polymerase binds to special genetic elements called promoter. The bacterial promoter is built with some genetic elements essential for efficient initiation of transcription (e.g. -10 and -30 blocks), thus we can switch on and off gene transcription by inserting or deleting some sequence blocks within promoter or even changing distance between them. This method of DNA replication or transcription control with use of automaton has one major advantages in comparison of natural scheme of control – it allows to make some logical calculations before cell take the final decision.

T. Krasiński received the PhD degree in mathematics from Institute of Mathematics of Polish Academy of Sciences in 1981, habilitation degree in 1992 from University of Łódź and professor degree in 2010 from the president of Poland. He is head of the Department of Algebraic Geometry and Theoretical Computer Science of University of Łódź. His interests are in analytic and algebraic geometry and recently in automata theory, formal languages and DNA computing.

S. Sakowski received the PhD degree in computer science from Silesian University of Technology in 2011, MSc in computer science from University of Łódź in 2004. His current interests are in DNA computing, computational biology, automata theory and formal languages.

T. Popławski received the PhD degree in biology from Institute of Microbiology University of Łódź in 2002, habilitation degree in 2011 from University of Łódź. He is a member of the Department of Molecular Genetics of University of Łódź. His primary interests are in DNA damage and repair and his secondary interests are in DNA computing.

# Appendix

Process of computing of the word *w*=*aabccc* by the push-down automaton from Example 2.

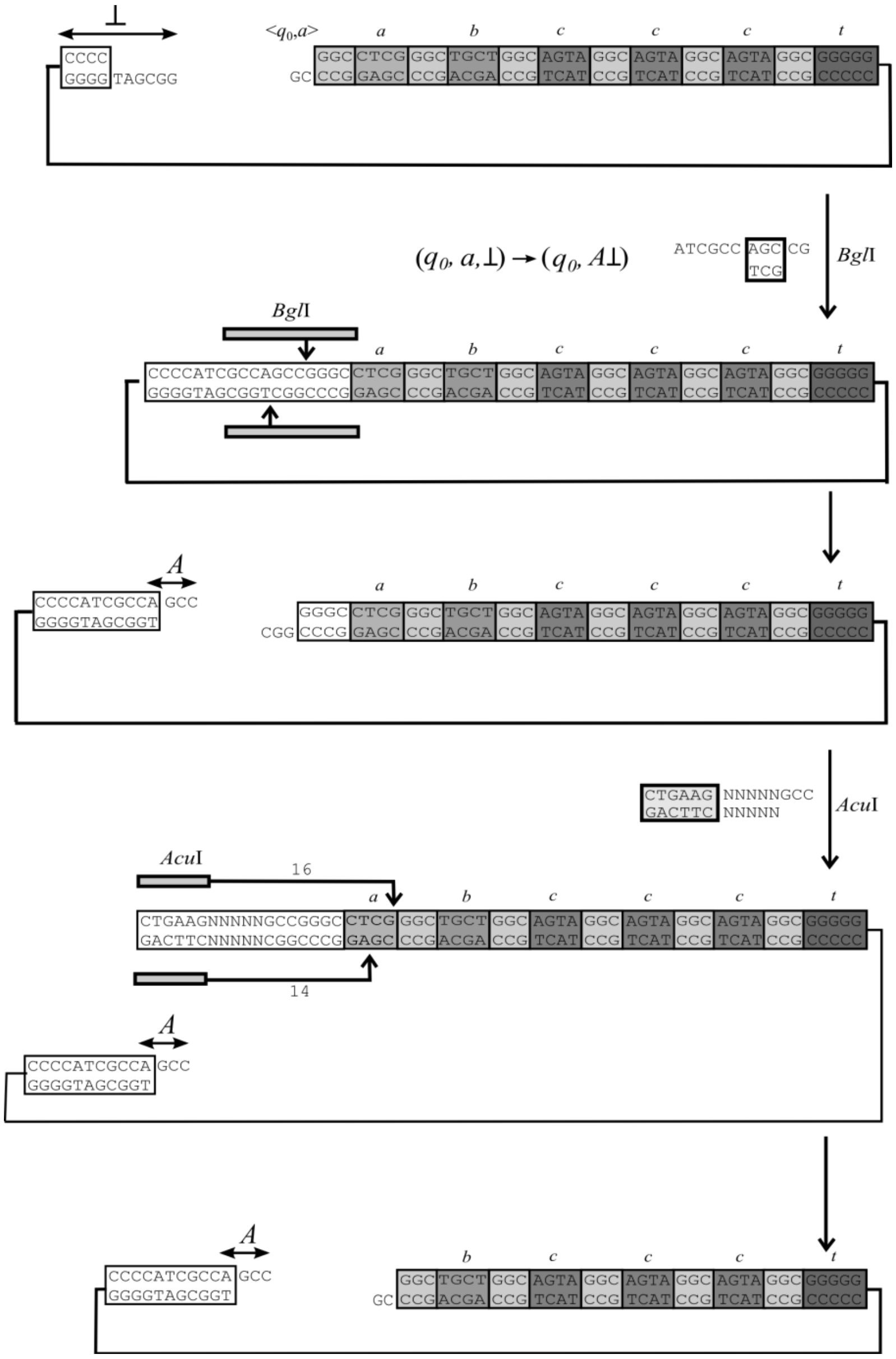

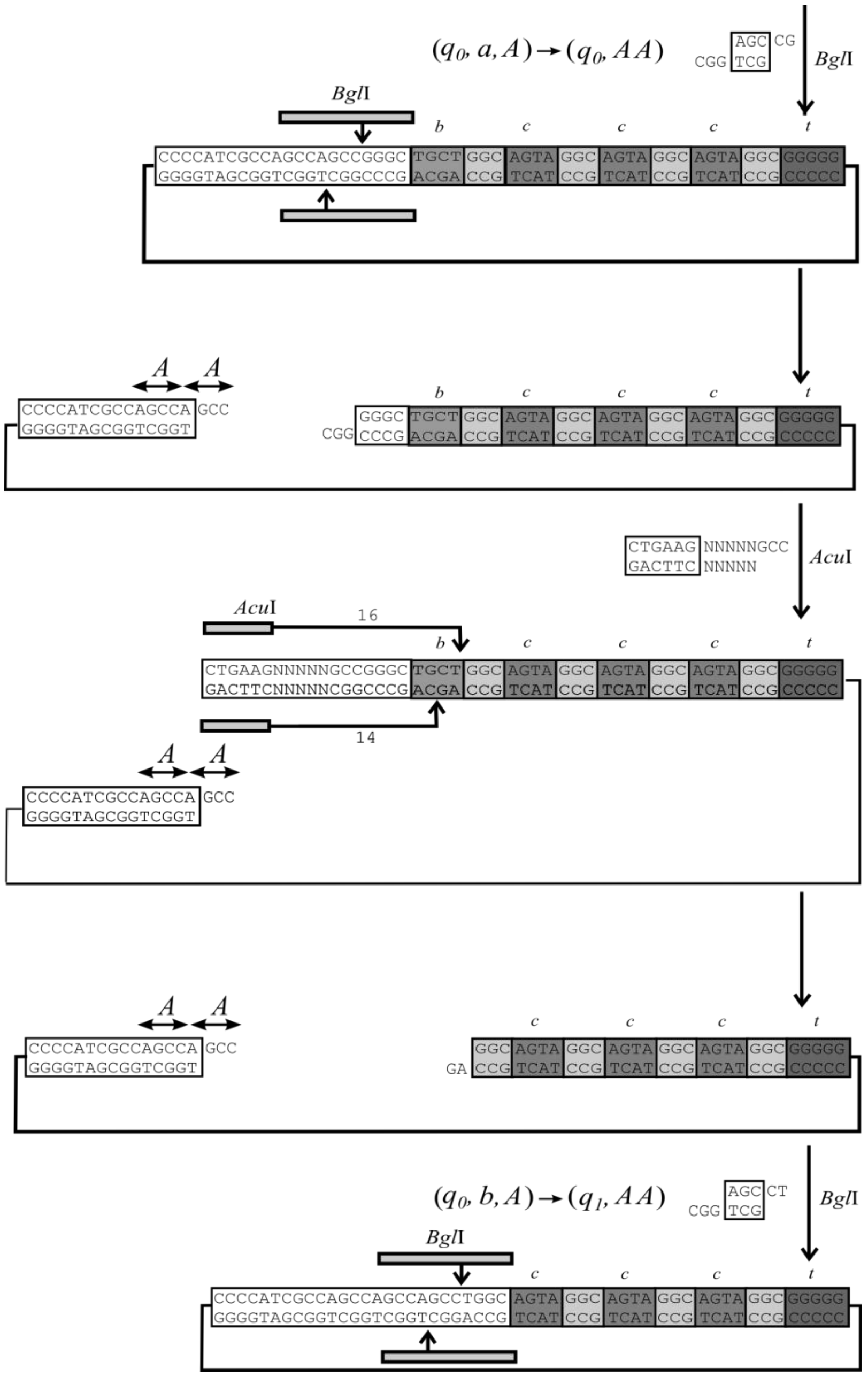

$(q_0, a, A) \rightarrow (q_0, AA)$
AGC CG
CGG TCG
BglI
BglI
b c c c t
CCCCATCGCCAGCCAGCCGGGC TGCT GGC AGTA GGC AGTA GGC AGTA GGC GGGGG
GGGGTAGCGGTCGGTCGGCCCG ACGA CCG TCAT CCG TCAT CCG TCAT CCG CCCCC
A A
CCCCATCGCCAGCCA GCC
GGGGTAGCGGTCGGT
b c c c t
GGGC TGCT GGC AGTA GGC AGTA GGC AGTA GGC GGGGG
CGG CCCG ACGA CCG TCAT CCG TCAT CCG TCAT CCG CCCCC
CTGAAG NNNNNGCC
GACTTC NNNNN
AcuI
AcuI
16
b c c c t
CTGAAGNNNNNGCCGGGC TGCT GGC AGTA GGC AGTA GGC AGTA GGC GGGGG
GACTTCNNNNNCGGCCCG ACGA CCG TCAT CCG TCAT CCG TCAT CCG CCCCC
14
A A
CCCCATCGCCAGCCA GCC
GGGGTAGCGGTCGGT
A A
CCCCATCGCCAGCCA GCC
GGGGTAGCGGTCGGT
c c c t
GGC AGTA GGC AGTA GGC AGTA GGC GGGGG
GA CCG TCAT CCG TCAT CCG TCAT CCG CCCCC
$(q_0, b, A) \rightarrow (q_1, AA)$
AGC CT
CGG TCG
BglI
BglI
c c c t
CCCCATCGCCAGCCAGCCAGCCTGGC AGTA GGC AGTA GGC AGTA GGC GGGGG
GGGGTAGCGGTCGGTCGGTCGGACCG TCAT CCG TCAT CCG TCAT CCG CCCCC

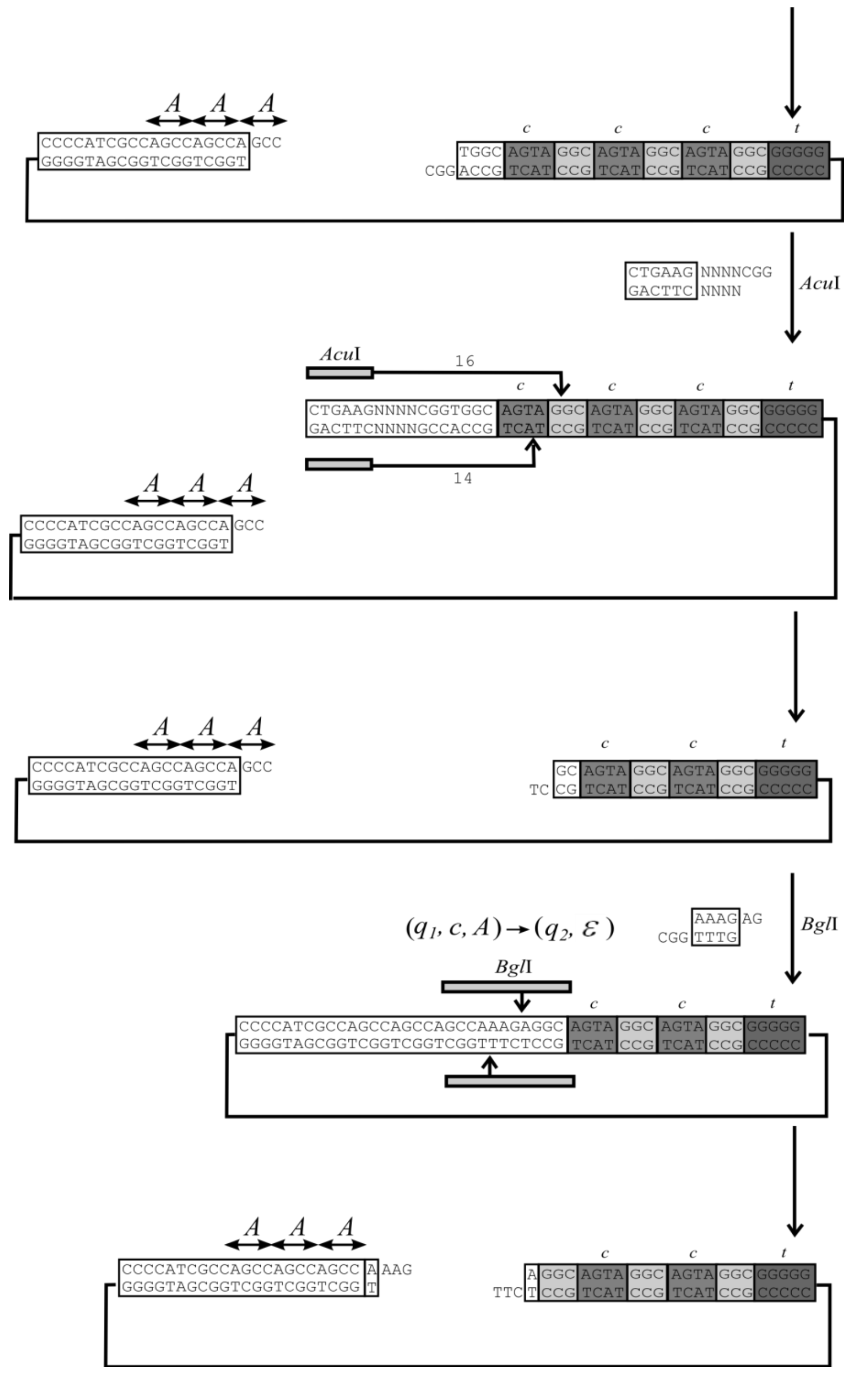

A A A
CCCCATCGCCAGCCAGCCA GCC
GGGGTAGCGGTCGGTCGGT
c c c t
TGGC AGTA GGC AGTA GGC AGTA GGC GGGGG
CGG ACCG TCAT CCG TCAT CCG TCAT CCG CCCCC
CTGAAG NNNNNCGG
GACTTC NNNN
AcuI
16
14
CTGAAGNNNNNCGGTGGC
GACTTCNNNNNGCCACCG
GC
TC CG
$(q_1, c, A) \rightarrow (q_2, \varepsilon)$
AAAG AG
CGG TTTG
BglI
CCCCATCGCCAGCCAGCCAGCCAAAGAGGC
GGGGTAGCGGTCGGTCGGTCGGTTTCTCCG
CCCCATCGCCAGCCAGCCAGCC A AAG
GGGGTAGCGGTCGGTCGGTCGG T
TTC

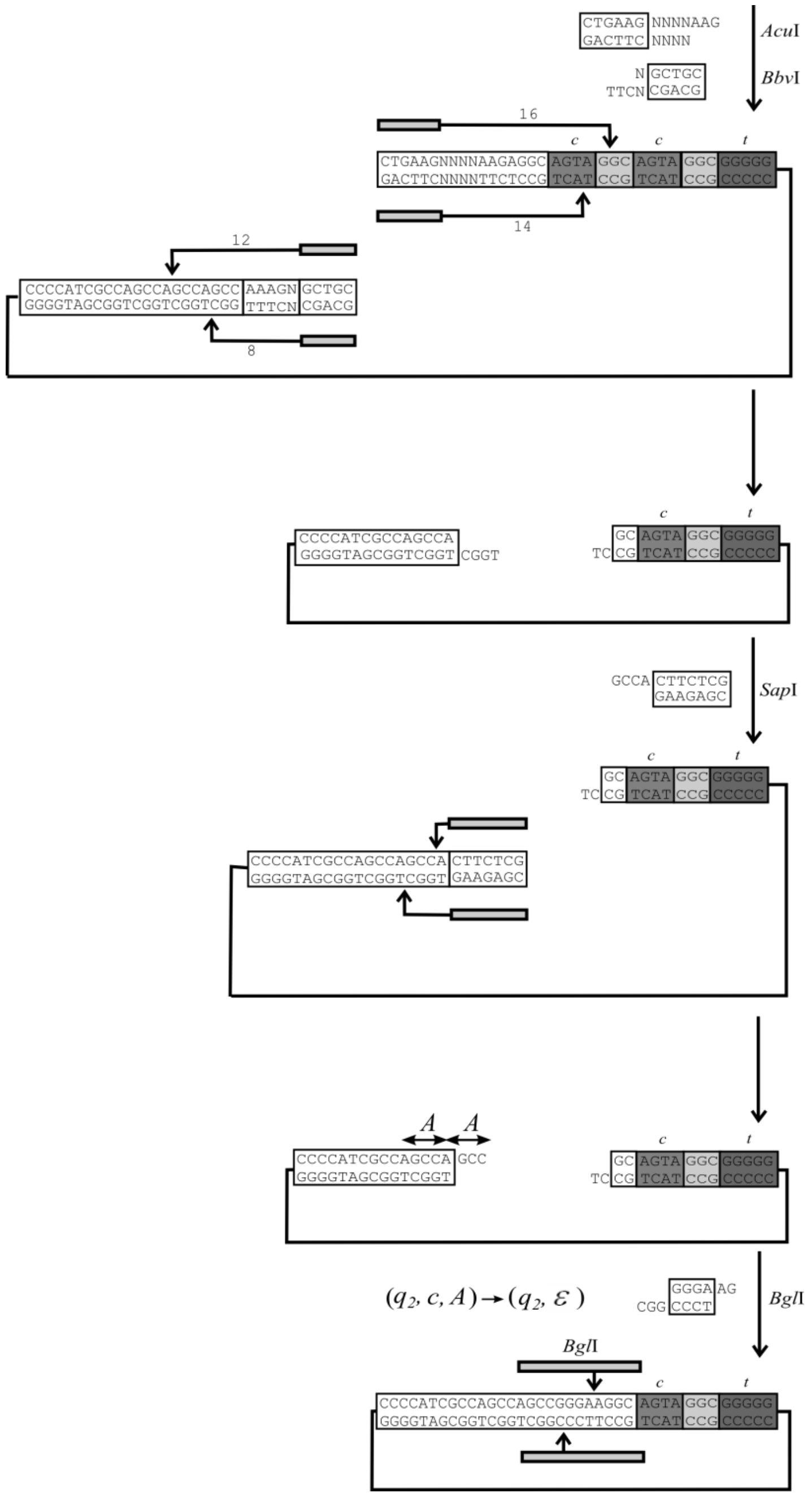
CTGAAG NNNNNAAG
GACTTC NNNN
AcuI
N GCTGC
TTCN CGACG
BbvI
16
c c t
CTGAAGNNNNNAAGAGGC AGTA GGC AGTA GGC GGGGG
GACTTCNNNNTTCTCCG TCAT CCG TCAT CCG CCCCC
14
12
CCCCATCGCCAGCCAGCCAGCC AAAGN GCTGC
GGGGTAGCGGTCGGTCGGTCGG TTTCN CGACG
8
c t
CCCCATCGCCAGCCA
GGGGTAGCGGTCGGT CGGT
GC AGTA GGC GGGGG
TC CG TCAT CCG CCCCC
GCCA CTTCTCG
GAAGAGC
SapI
c t
GC AGTA GGC GGGGG
TC CG TCAT CCG CCCCC
CCCCATCGCCAGCCAGCCA CTTCTCG
GGGGTAGCGGTCGGTCGGT GAAGAGC
A A
CCCCATCGCCAGCCA GCC
GGGGTAGCGGTCGGT
c t
GC AGTA GGC GGGGG
TC CG TCAT CCG CCCCC
(q2, c, A) → (q2, ε)
GGGA AG
CGG CCCT
BglI
BglI
c t
CCCCATCGCCAGCCAGCCGGGAAGGC AGTA GGC GGGGG
GGGGTAGCGGTCGGTCGGCCCTTCCG TCAT CCG CCCCC

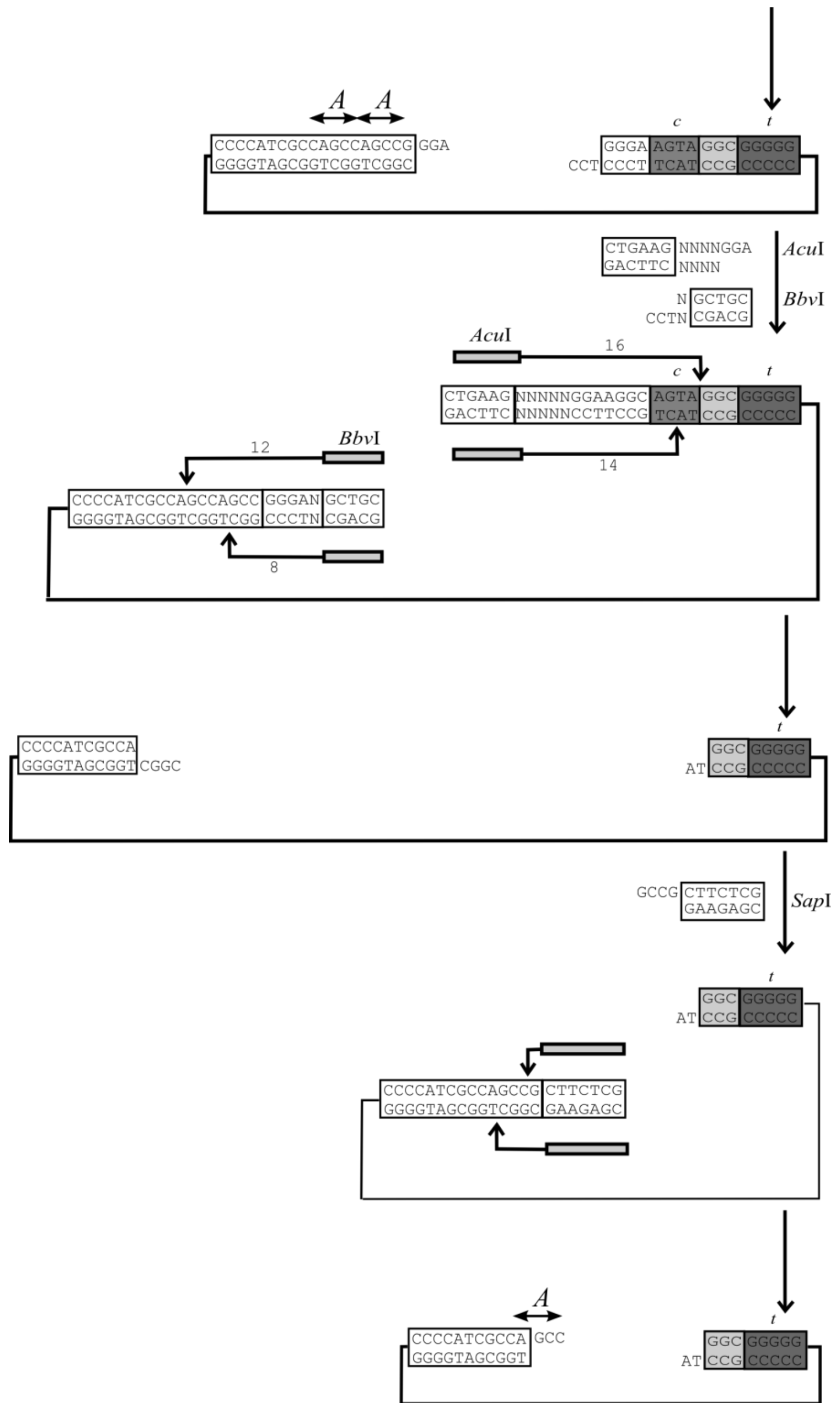

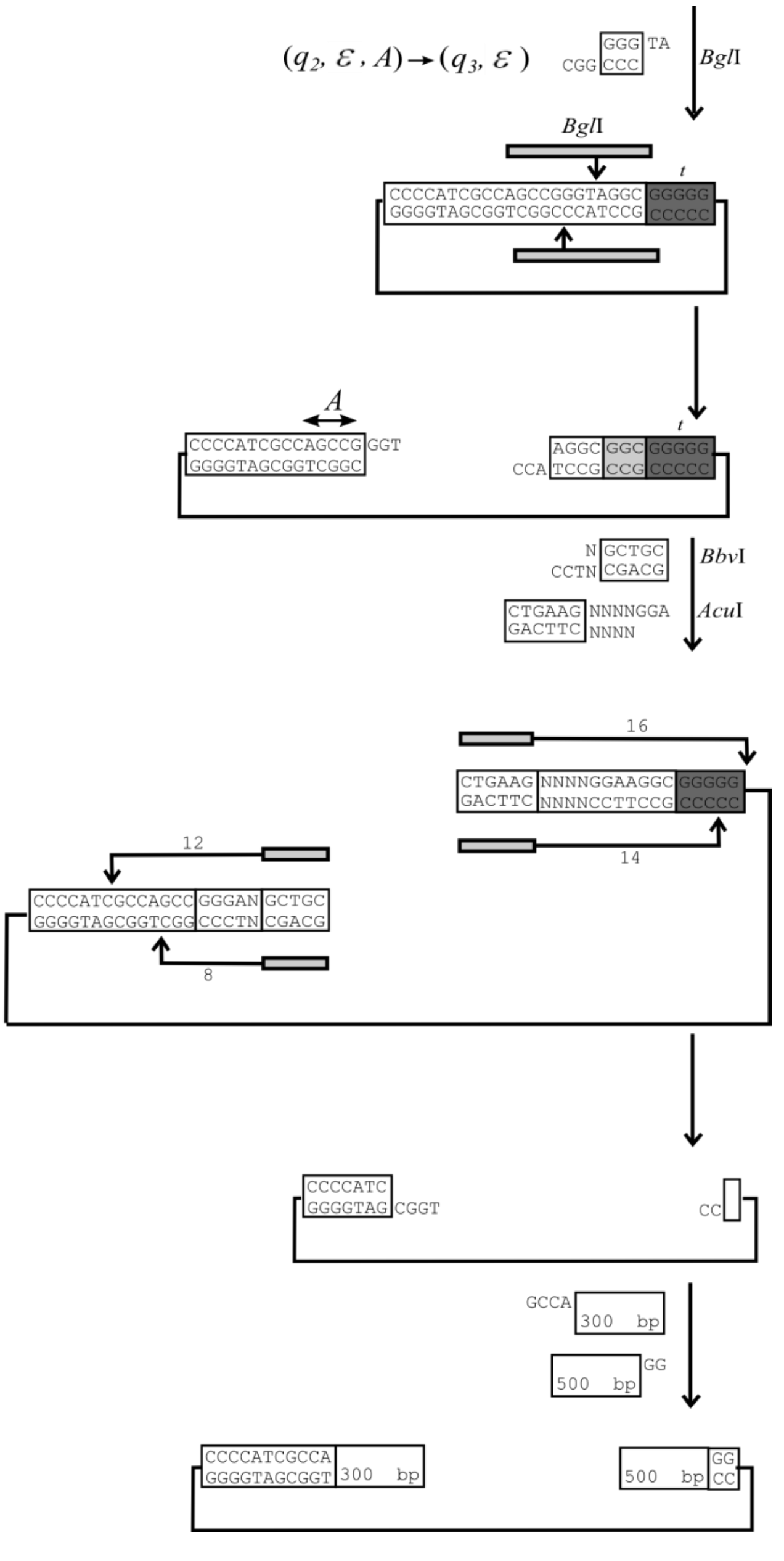